\documentclass[twocolumn]{article}
\usepackage[dvips]{graphicx,color}
\newcommand{\eq}[1]{\label{#1} }

\title{On the X-ray Image of The Crab Nebula: Comparison with Chandra Observations}

\author{Shinpei SHIBATA $^{1}$, Haruhiko TOMATSURI$^{1}$, \\ Makiko SHIMANUKI$^{1}$, Kazuyuki SAITO$^{1}$, Koji MORI$^{2}$ \\ $^{1}$Department of Physics Yamagata University, Yamagata 990-8560, JAPAN, \\ $^{2}$Department of Astronomy and Astrophysics, 525 Davey Laboratory, \\ The Pennsylvania State University, University Park, PA 16802, USA}

\newcommand{\bmath}[1]{\mbox{\boldmath $#1$}}
\begin{document}
\date{in original form 2003 June 23, revised 5 August, accepted 25 August 2003 for MNRAS}

\maketitle

\label{firstpage}

\begin{abstract}

An axisymmetric model for the Crab Nebula is constructed to examine the
flow dynamics in the nebula.  The model is based on that of Kennel and
Coroniti (1984), although we assume that the kinetic-energy-dominant
wind is confined in an equatorial region. The evolution of the
distribution function of the electron-positron plasma flowing out in the
nebula is calculated.  Given viewing angles, we reproduce an image of
the nebula and compare it with Chandra observation.

The reproduced image is not a ring-like but rather 'lip-shaped'.  It is
found that the assumption of toroidal field does not reproduce the
Chandra image. We must assume that there is disordered magnetic field
with an amplitude as large as the mean toroidal field. In addition, the
brightness contrast between the front and back sides of the ring cannot
be reproduced if we assume that the magnetization parameter $\sigma$ is
as small as $\sim 10^{-3}$.  The brightness profile along the semi-major
axis of the torus is also examined. The non-dissipative, ideal-MHD
approximation in the nebula appears to break down.

We speculate that if the magnetic energy is released by some process
that produce turbulent field in the nebula flow and causes heating and
acceleration, e.g. by magnetic reconnection, then the present
difficulties may be resolved (i.e. we can reproduce a ring image, and a
higher brightness contrast). Thus, the magnetization parameter $\sigma$
can be larger than previously expected.

\end{abstract}

stars: pulsars: general -- ISM: individual: Crab Nebula

\section{Introduction}

A standard picture of the Crab Nebula was given by Kennel and Coroniti
(KC; 1984).  According to their picture, a super-fast
magnetohydrodynamic wind, which is generated by the central pulsar,
terminates at a shock, with the nebula identified as a postshock flow
shining in synchrotron radiation.  The central cavity of the nebula is
occupied by the unseen wind.  The shock is supposed to occur at the
standing inner wisp.

The KC model is very successful explaining the synchrotron luminosity,
spectrum and frequency-dependent size of the nebula.  An important
conclusion of the KC model is that the energy of the wind is conveyed
mostly by kinetic energy in the bulk motion of the plasma.  Because the
energy flux is in the form of an electromagnetic field at the base of
the wind, this means that the efficiency of the wind acceleration is
extremely high; KC found it to be 99.7\%.

The principal parameters of the pulsar wind are (1) its luminosity
$L_w$, (2) the Lorentz factor $\gamma_w$ of the bulk flow and (3) the
ratio $\sigma$ of the electromagnetic energy flux to the kinetic energy
flux, which is referred to as the magnetization parameter.  $L_w$ is
essentially the spin-down luminosity $\approx 5 \times 10^{38}$ erg
s$^{-1}$.  The remaining two parameters, $\gamma_w$ and $\sigma$,
together with the nebula pressure $P_N$, or equivalently the
equipartition field $B_{eq} = \sqrt{4 \pi P_N}$, determine the overall
synchrotron spectrum.  Conversely, the synchrotron spectrum tells us
about the parameters.  Given the synchrotron luminosity of $2 \times
10^{37}$erg s$^{-1}$, the nebula size of $\sim 1$~pc, and the peak and
turn-off synchrotron spectrum energies of 2~eV and $10^8$~eV,
respectively, one finds $\gamma_w =3.3 \times 10^6$, $\sigma = 3.8
\times 10^{-3}$, and $B_{eq} = 0.38$ mG. This result can be obtained
even with an order-of-magnitude estimate (Shibata, Kawai and Tamura
1998). More rigorous fitting of the observed spectrum of the whole
nebula gave similar values (e.g., KC, Atoyan and Aharonian 1996).  The
field strength has been confirmed by observations of inverse Compton
emission in the TeV band (Weekes et al. 1989; Hillas et al. 1998).
Thus, the dominance of the kinetic energy of the wind seems very firm.

From a theoretical point of view, however, the smallness of $\sigma$, or
in other words dominance of the kinetic energy, is a mystery. No wind
theory has been able to explain how such a high efficiency of
acceleration is achieved.

Chandra observation clearly shows a disk-jet structure and moving wisps
with seeds of $\sim 0.45 c$ (Mori 2002), where $c$ is the speed of
light.  Because the KC model is spherically symmetric and steady, it may
seem insufficient to understand the highly structured and dynamical
nature seen by Chandra.  However, the basic idea that a kinetic dominant
wind shocks and shines seems still firm and convincing.  One may assume
that the equatorial wind has different parameters than the polar wind.
Such a latitude dependence of the wind parameters may suffice to explain
the apparent disk-jet structure although how such a latitude dependence
is made is not known.

In this paper, we suggest that high spatial resolution of Chandra
affords a chance to examine the assumptions which were made in the KC model
but have yet to be checked. Among the assumptions, the ideal-MHD
condition (no dissipation) and toroidal field approximation are of
particular importance.  If these assumptions are not adequate, the past
conclusion of small $\sigma$ may need to be reconsidered.

We model the nebula in 3-dimensions based on the KC picture and
reproduce an image, which can be compared with the Chandra observation
(Mori 2002) is made.  We shall show that a considerable change to the KC
picture is required to reproduce the Chandra image.  In this paper, we
suggest that disordered magnetic field in the nebula is needed . Some
process which converts magnetic energy into thermal and kinetic energy,
such as magnetic reconnection, may take place in the nebula.  In a
subsequent paper, spatially-resolved X-ray spectra will be described and
compared with Chandra results.

\section{A 3D Model}

We postulate that the nebula flow obeys the KC steady solution and do
not solve the dynamics. The properties of the KC flow are summarized as
follows.  If $\sigma$ is much less than unity as was suggested, the
speed of the flow is $\sim (1/3)c$ just after the shock and decreases
rapidly with distance from the pulsar $R$ as $V \propto R^{-2}$; because
the flow is subsonic, the pressure and density $n$ are roughly uniform
as for adiabatic expansion such that the mass conservation, $nR^2V
\approx $const.(implying a decrease of the flow velocity).  Due to
deceleration, the magnetic field accumulates and is amplified according
to the frozen-in condition, $B \propto r$.  Once the magnetic field
grows as large as the equipartition field, the magnetic pressure becomes
important in the flow dynamics.  As a result, the flow speed saturates.
This takes place where the nebula is brightest (at $\sim
(3\sqrt{\sigma})^{-1}$ shock radii).  The smaller $\sigma$, the larger
and brighter the nebula. The smallness of $\sigma$ is then required to
explain the luminosity and the extent of the nebula.  The indicated flow
speed is small if $\sigma$ is small: $V/c \sim 3 \sigma$.  It is notable
here that the above flow dynamics depends on the assumption of the
ideal-MHD condition.

The KC model is spherically symmetric and obviously inadequate to
account for the observed morphology.  We therefore restricted ourselves
to an equatorial region of the KC spherical model with half width of
$\Theta_{eq} \sim 10^\circ$ and cut the intermediate latitude regions
out, such that the disk may be reproduced in an image (see Fig.~1).
Although we assume a polar flow by leaving the polar region with a
semi-opening angle of $\Theta_{pol} \sim 10^\circ$ for reproduction of
the jet image, this is just in an artist's spirit, and we do not provide
any analysis of the polar jets in this paper.

\begin{figure}
  \begin{center}
    \includegraphics[height=13pc]{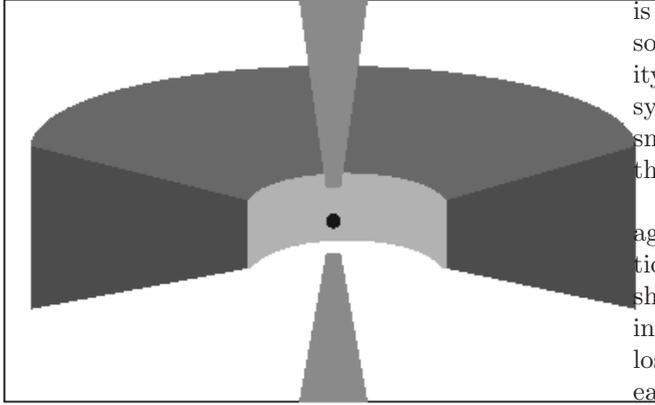}
  \end{center}
  \caption{The three-dimensional structure we assumed for reproduction of image.
The spherical flow by Kennel and Coroniti (1984) is picked up for the disk 
and polar flows.}
\end{figure}

Our kinematic scheme to reproduce images of the nebula is made so that
different types of the flow dynamics can be applied in the future; i.e.,
for a given velocity field, we trace a fluid element and associated
particle distribution function according to the Lagrangian view of
fluid.

Given the radial velocity field $V(t,R)$ as a function of time and
radial distance from the pulsar, the position of the fluid element, the
toroidal magnetic field, and the proper density are respectively
obtained from
\begin{eqnarray}
{D  R \over Dt}             & = & V , \eq{rv}
\\
{D  \over Dt} \left( \ln B \right)       & = & 
- \left( {V \over R} + {\partial V \over \partial R } \right) ,
\\
{D  \over Dt} \left( \ln n \right)       & = & 
 - {D  \over Dt} \left( \ln \Gamma \right) 
 - \left( { 2 V \over R } + { \partial V \over \partial R } \right) ,
\end{eqnarray} 
where $t$ is the observer's time and $\Gamma = (1-V^2/c^2)^{-1/2}$ is
the Lorentz factor of the flow.  We use the KC solution, which is given
analytically, for the velocity field.  Although the KC solution does not
include synchrotron losses, its effect on $V$ is supposed to be small
because synchrotron losses are about 10\% of the kinetic energy of the
flow.

For the energy distribution of the particles, we again invoke the KC
picture: a power law distribution is built up immediately after the
shock, and the shock-accelerated particles simply lose their energy in
the postshock flow by adiabatic and synchrotron losses.  We trace the
energy $\epsilon$ (normalized by $mc^2$) of each particle in a fluid
frame by
\begin{equation}
{D  \over Dt^\prime} \left( \ln \epsilon \right)        =  
{1 \over 3} {D  \over Dt^\prime} \left( \ln n \right) 
+ {1 \over \epsilon} \left( d \epsilon \over d t^\prime \right)_{\rm loss} ,
\eq{fele} 
\end{equation}
where
\begin{equation}
- \left( d \epsilon \over d t^\prime \right)_{\rm loss} =
{ 4 \over 3} \sigma_{\rm T} c \epsilon^2 U_{\rm mag} ,
\end{equation}
and the magnetic energy density, $U_{\rm mag} = B^2/8 \pi \Gamma^2$, is
measured in the proper frame, and $\sigma_{\rm T}$ is the Thomson cross
section.  The proper time $t^\prime$ is related to $t$ by $D t /
Dt^\prime = \Gamma$.

The distribution function is defined by
\begin{equation} \eq{deff}
dn = f(t, R; \epsilon , \theta) \sin \theta d \theta d \varphi d \epsilon, 
\end{equation}
where $\theta$ is the pitch angle with respect to the local field and
$\varphi$ is the azimuth. Note that the distribution function is defined
in the flow proper frame.  We assume that the postshock distribution
follows a power law with index $p$ ($\geq 1$) in between the minimum
energy $\epsilon_{\rm min}$ and the maximum energy $\epsilon_{\rm max}$,
and it is isotropic such that
\begin{equation} \eq{defff}
f_{\rm i} (\epsilon_{\rm i} )
= {K \over 4 \pi } n_{\rm i} \epsilon_{\rm i}^{-p},
\end{equation}
where the suffix `i' indicates `injection' at the postshock region,
and the normalization is given by 
\begin{equation}
K= \left\{
\begin{array}{cc}
{\displaystyle { (p-1) \epsilon_{\rm min}^{(p-1)} \over 
1 - (\epsilon_{\rm min} / \epsilon_{\rm max} )^{p-1} }
}
&
p \neq 1
\\
{\displaystyle 
{ 1 \over \ln \left( \epsilon_{\rm max} / \epsilon_{\rm min} \right)}
}
& 
p=1
\end{array}
\right.
\end{equation}
and $n_{\rm i} = \int\!\!\int\!\!\int f_{\rm i} ( \epsilon ) \sin \theta
d \theta d \varphi d \epsilon$ gives the postshock proper density.
$\epsilon_{\rm max}$ is assumed to be a maximum attainable value, $e B_2
R_{\rm s} / mc^2$, where $R_{\rm s}$ is the shock distance from the
pulsar, and $B_2$ is the postshock field.  $\epsilon_{\rm min}$ is
determined so that the pressure calculated from $f_{\rm i}$ satisfies
the shock jump condition.

We solve (\ref{fele}) numerically for a sample of particles
in a given fluid element, and thereby we obtain $\epsilon$ 
as a function of $\epsilon_{\rm i}$ and $t$.
Then we calculate distribution functions from
\begin{equation} \eq{evolf}
f(\epsilon (R))= 
{n \over n_{\rm i} }
f_{\rm i} (\epsilon_{\rm i})
{d \epsilon_{\rm i} \over d \epsilon } .
\end{equation}
For steady state models, integration for a single
fluid element gives distribution functions in the whole nebula.

It is obvious that the above kinematic scheme can be easily generalized
for non-steady and non-radial flow, which can be obtained by numerical
MHD simulations.

\section{Reproduction of the Nebula Image}
\label{imageapp}

\subsection{Synchrotron Specific Emissivity}

Once the evolution equations (\ref{rv})-(\ref{fele}) are solved, and the
distribution function is obtained by (\ref{evolf}), it is straight
forward to get volume emissivity, which is integrated to give a nebula
image.

Since the nebula flow is relativistic, a Lorentz transformation is
applied between the flow proper frame and the observer's frame (or
rather the pulsar frame, in which the pulsar is at rest). Let us denote
the 4-vector of a synchrotron photon by ($\omega / c$, $\bmath{k}$) in
the observer's frame and ($\omega^\prime / c$, $\bmath{k}^\prime$) in
the proper frame.  If the ideal-MHD condition $\bmath{E} + \bmath{V}
\times \bmath{B}/c = 0$ holds in the nebula flow, the transformation of
the electromagnetic field is simpler:
\begin{eqnarray}
\bmath{E}^\prime  &  =  &  0, \\
\bmath{B}^\prime_\parallel & = & \bmath{B}_\parallel 
= (\bmath{V} \cdot \bmath{B} ) \bmath{V} / V^2, \\
\bmath{B}^\prime_\perp & = & \bmath{B}_\perp / \Gamma
= (\bmath{B} - \bmath{B}_\parallel ) / \Gamma,
\end{eqnarray}
where the primes indicate the quantities in the flow frame and
$\parallel$ and $\perp$ are based on the directed of the flow velocity
$\bmath{V}$.  There is no electric field in the plasma flow frame.

The spectral power of a relativistic particle with pitch angle $\theta$
(the angle of the particle motion to the local magnetic field in the
proper frame) is
\begin{equation} \eq{Pows1}
{\cal P}_{\rm s1} ( \omega^\prime, \epsilon, \theta ) = 2 \sigma_{\rm T} c
U_{\rm mag} \epsilon^2 \sin^2 \theta \ {\cal S}(\omega^\prime ;
\omega_{\rm c}).
\end{equation}
For the monochromatic approximation, we use
${\cal S}(\omega^\prime ; \omega_{\rm c}) = \delta (\omega^\prime -
\omega_{\rm c})$, and for the relativistic approximation,
\begin{equation}
{\cal S}(\omega^\prime ; \omega_{\rm c}) = 
{9 \sqrt{3} \over 8 \pi \omega_{\rm c} }
F \left( \omega^\prime \over \omega_{\rm c} \right)
\end{equation}
where
$\displaystyle F(x)=x \int^\infty_x K_{5 \over 3} (\xi ) d\xi$, and
\begin{equation}
\omega_{\rm c} =
{3 e |\bmath{B}^\prime | \epsilon^2 \sin \theta \over 2 m c }
\end{equation}
is the critical frequency.  The synchrotron power of a single particle
is strongly beamed within a width of $\sim \epsilon^{-1}$.  Therefore,
the emission into a frequency interval $d \omega^\prime$ and in a solid
angle $d \Omega^\prime$ directed toward the observer is given by
\begin{equation} \eq{jomega}
j_{\omega^\prime} (\theta ) d \omega^\prime  d \Omega^\prime  =
\int_0^\infty {\cal P}_{\rm s 1} (\omega^\prime, \theta, \epsilon)
f(\epsilon, \theta) d \epsilon d \Omega^\prime d \omega^\prime ,
\end{equation}
where $\theta$ is given by $\cos \theta = \bmath{n}^\prime \cdot
\bmath{B}^\prime / |\bmath{B}^\prime |$, and $\bmath{n}^\prime$ is the
unit vector directing to the observer in the proper frame. Below,
$\bmath{n}$ indicates the observer's direction in the observer's frame.


For the link between the proper frame and the observer's frame,
we include the Doppler effects,
\begin{equation} \eq{dopper}
\omega = { \omega^\prime \over \Gamma (1 - \beta \mu) }  \ \ and \ \ 
\mu = { \mu^\prime + \beta \over 1 + \beta \mu^\prime },
\end{equation}
where $\mu = \bmath{n} \cdot \hat{\bmath{V}}$ and $\mu^\prime =
\bmath{n}^\prime \cdot \hat{\bmath{V}}$.  The unit vector of the flow
direction is denoted by $\hat{\bmath{V}}$.  The transformation between
the received power $d P_r$ and the emitted power $d P^\prime$ (Rybicki
\& Lightman 1979) is given by
\begin{equation}
{d P_r \over d \Omega d \omega} =
\Gamma^3 (1 + \beta \mu^\prime )^3
{d P^\prime \over d \Omega^\prime d \omega^\prime} =
{1 \over \Gamma^3 (1 - \beta \mu )^3}
{d P^\prime \over d \Omega^\prime d \omega^\prime}
\end{equation}
where the Doppler effect (\ref{dopper}) has been taken into account.
Finally the Lorentz contraction is $dN = \Gamma f(\epsilon, \theta ) d
\Omega^\prime d \epsilon$, where $N$ is the number density in the
observer's frame.  Thus the emissivity in the observer's frame becomes
\begin{equation}
j_\omega( \bmath{n} ) =
C \int {\cal P}_{\rm s1} (\omega^\prime, \theta, \epsilon) 
f(\epsilon, \theta)d \epsilon,
\end{equation}
where 
\begin{equation}
C = \Gamma^4 ( 1+\beta \mu^\prime )^3=
{1 \over \Gamma^2(1 - \beta \mu)^3} .
\end{equation}

\subsection{Viewing Angle}

In order to specify the viewing angle of the observer, we relate the
`observer's coordinate' $\bmath{X}=$ ($X$, $Y$, $Z$), where $+X$
directed toward the observer and $+Z$ directed toward north on the sky,
to the `nebula coordinate' $\bmath{x}=$ ($x$, $y$, $z$), where the
$z$-axis coincides with the symmetry axis of the nebula which is
believed to be the rotation axis of the pulsar.  We use 48$^\circ$ and
28$^\circ$ as the position angle and the inclination angle of the
symmetry axis, respectively.

An image of the nebula is obtained by
\begin{equation} \eq{imagint}
I_\omega (Y,Z) = \int_{- \infty}^\infty 
j_\omega (X, Y, Z, \bmath{n}) \; d X.
\end{equation}

\subsection{Summary of the procedure}

For a given observation frequency, the integration (\ref{imagint}) is
done numerically for each `pixel' at ($Y$, $Z$).  The integrand is
calculated as follows:
\begin{enumerate}
\item Given $\bmath{X}=(X, Y, Z)$ and $\omega$,
\item transformation from the observer's coordinates to 
the nebula coordinates is done; 
for the position, $(X,Y,Z) \ \rightarrow \ (x,y,z)$, and
also for the  componets of the observer's direction,
$\bmath{n}=(1,0,0) \ \rightarrow \ \bmath{n}=(n_x,n_y,n_z)$.

\item 
The flow velocity $\bmath{V} (R)$ and the magnetic field $\bmath{B} (R)$
at the point
are obtained by using the KC solution. Note that the velocity and the magnetic
field are respectively radial and azimuthal in the 'nebula coordinate'.

\item The observation frequency, the direction and the local magnetic field
are transformed into those in the flow frame:
$\omega \rightarrow \omega^\prime$,
$\bmath{n} \rightarrow \bmath{n}^\prime$,
$\bmath{B} \rightarrow \bmath{B}^\prime$.

\item The pitch angle of the particles directed toward the
observer is obtained.

\item Regarding the distribution function at ($x$, $y$, $z$), the
emissivity in the flow frame is calculated 
(we use the monochromatic approximation).

\item Finally, the emissivity is converted to the volume emissivity $j_\omega
(\bmath{n})$ at ($X$, $Y$, $Z$) in the observer's frame by multiplying
by the Doppler factor $C$.

\end{enumerate}

\section{Results}

One may expect a ring-like structure in the reproduced image, such as
observed with Chandra (see the top panel of Fig.~2), since we have
assumed that the flow is restricted within a disk.  The expected radius
of the ring will be $\sim 1/3\sqrt{\sigma}$ (about 6 shock radii for
$\sigma = 0.003$), at which point the nebula brightens due to the
amplified magnetic field.  However, what we have is not a ring-like but
is rather a 'lip-shaped' image shown in the bottom panel of Fig.~2.  At
the north-east and south-west corners of the expected ellipse (ring),
the pitch angles of the particles directed toward us are small, and
therefore the surface brightness is reduced. This effect combined with
the central cavity yields an image which is 'lip-shaped'. The smallness
of pitch angle actually has two effects. One is that the single-particle
emissivity is proportional to $\sin^2 \theta$, which is small. The other
is due to number of contributing particles. For a given observation
frequency, the energy of the particles radiating at the frequency is
higher for smaller pitch angles, so that the number of particles
contributing to the frequency is smaller because of the negative slope
of the distribution function.

\begin{figure}
  \begin{center}
    \includegraphics[height=5cm]{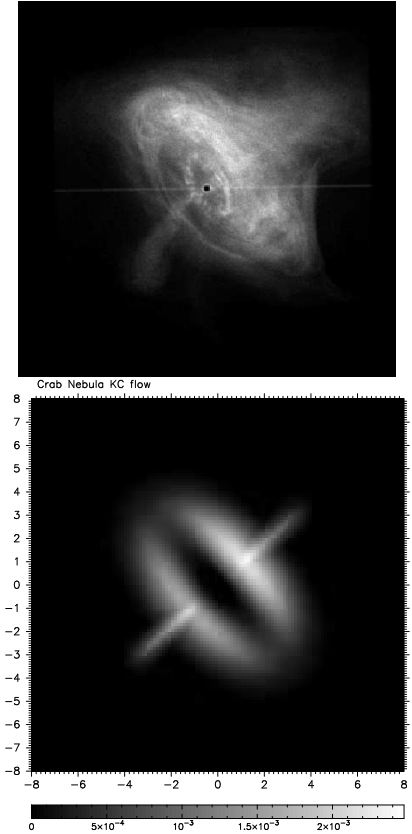} \\
    \includegraphics[height=6.0cm]{fig2b.ps}
  \end{center}
  \caption{The Chandra image (top) and a reproduced image (bottom),
  where we assume a postshock flow with $\sigma = 0.003$ by Kennel and
  Coroniti (1984) but the flow is assumed to be restricted within an
  equatorial region.  For the bottom image, the gray scale is in units
  of 0.016 erg s$^{-1}$ cm$^{-2}$ str eV.  }
\end{figure}

Another important point to consider is the intensity ratio between the
front and back sides of the ring.  We obtain a value of 1.3, but
observed value is $\sim 5$ (Pelling et al. 1987, Willingale et al., 2001).
The weak contrast is caused by deceleration of the nebula flow (i.e. the
smallness of $\sigma$).  Mori et al. (2003) suggest that the ratio is
about 3 with Chandra.  This value is still incompatible with the KC
picture.  As long as the intensity contrast is attributed to Doppler
boosting, such a weak contrast is unavoidable in the frame work of the
KC model.

\begin{figure}
  \begin{center}
    \hspace*{3mm} 
    \includegraphics[width=14pc]{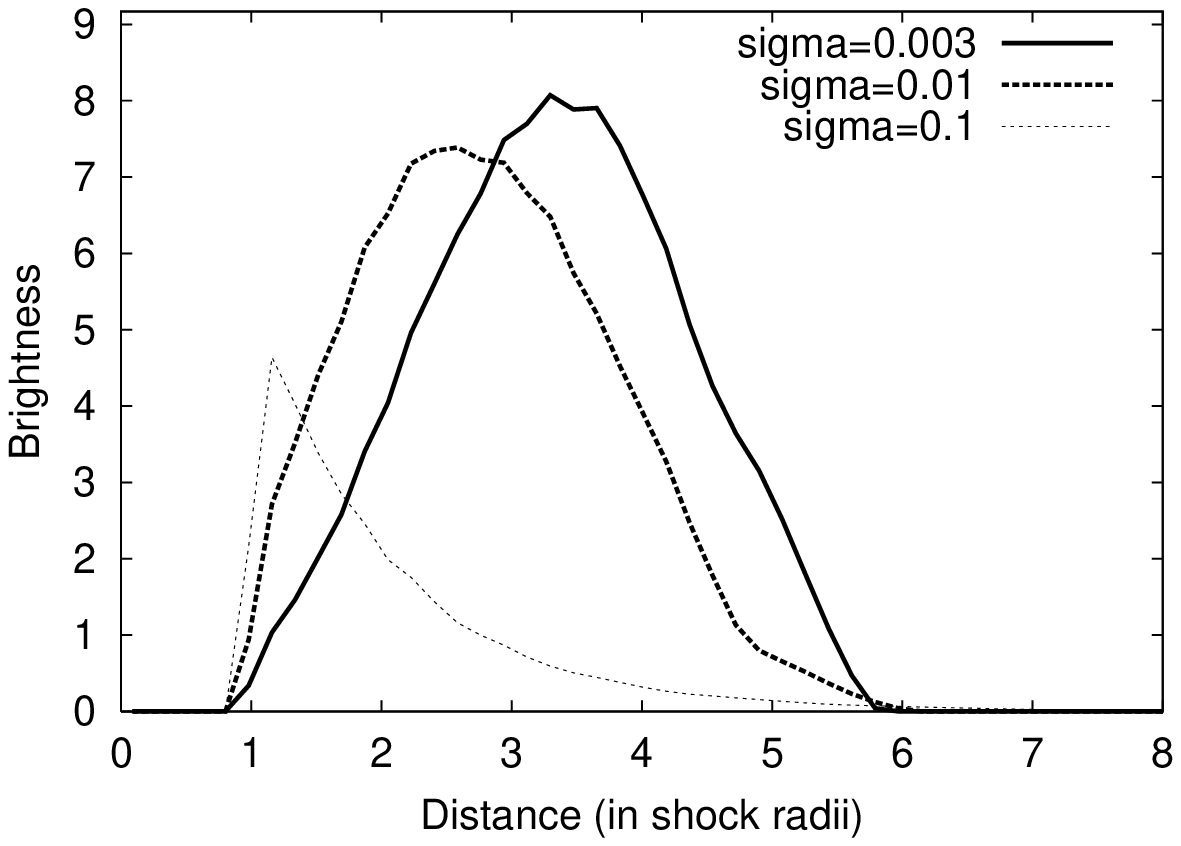} \\
    \includegraphics[width=14pc]{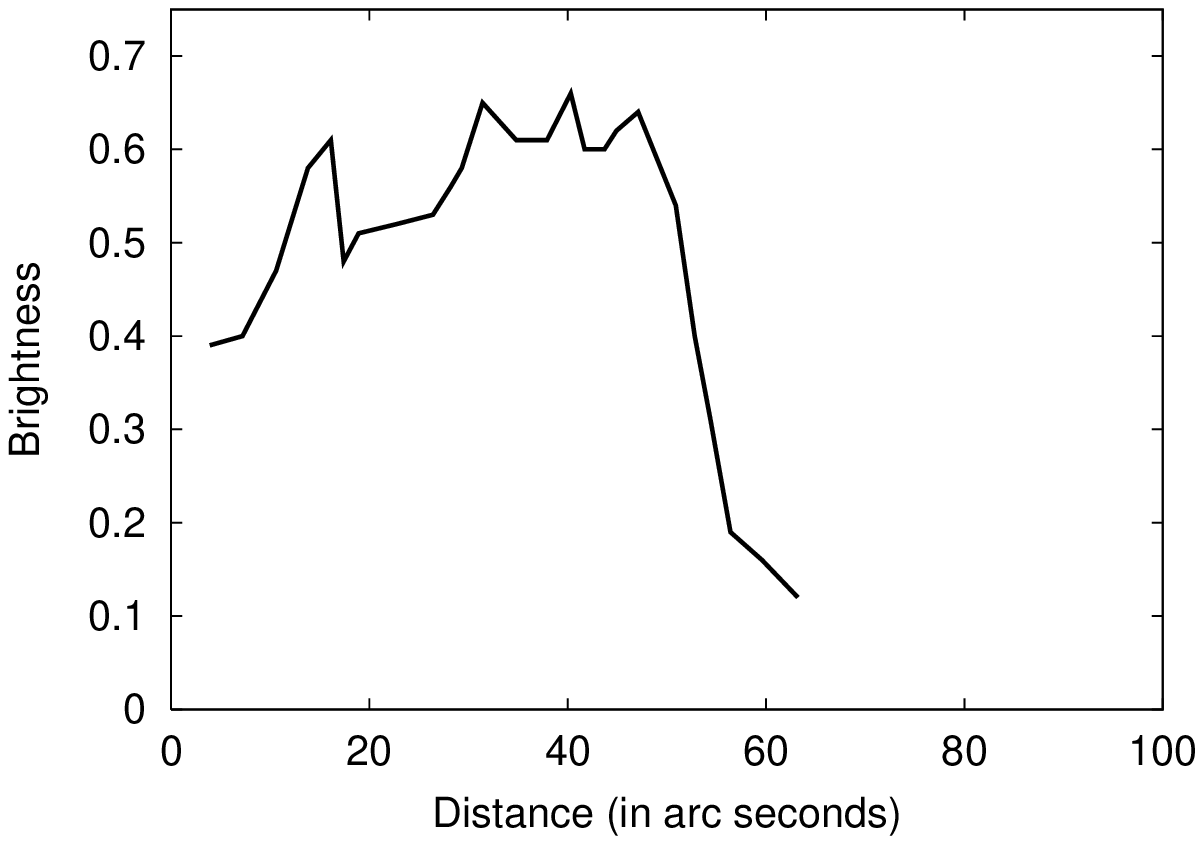}
  \end{center}
  \caption{Surface brightness along the semi-major axis of the torus in
  the north-east direction. The upper panel shows the calculation with
  different $\sigma$'s while the lower panel is measured from the
  Chandra observation (Mori 2002).  The distance is in units of the
  shock radius (which is supposed to be about $13^{\prime
  \prime}-14^{\prime \prime}$) for the model, and in arc seconds for the
  observation. The brightness is in units of $10^6$ erg s$^{-1}$
  cm$^{-2}$ str$^{-1}$ eV$^{-1}$ for the model and in counts s$^{-1}$
  arcsec$^{-2}$ for the observation.  }
\end{figure}

How surface brightness changes with distance from the shock depends on
$\sigma$. Mori (2002) also measured the surface brightness along the
semi-major axis of the torus, from which Doppler boosting should not
affect the brightness.  This result (bottom panel of Fig.~3) is compared
with the present model (top panel of Fig.~3), for which we provide
curves of varying $\sigma$.  The present model does not reproduce the
first peak in the observation, which corresponds to the inner
ring. However, it is notable that the brightness distribution of the
inner ring is similar to that for the $\sigma = 0.1$ model.  The
location of synchrotron burn off is reproduced by the $\sigma =0.01$
model. Finally, we point out that the surface brightness decreases much
faster with distance in the observation than in the model.


As indicated by the 'lip-shaped' image, the absolute value of the
surface brightness is much less than observed along the semi-major axis.
Because the reproduced image includes only the disk component, to which
we restricted ourselves (rather than assuming the spherical KC model),
the X-ray luminosity of the reproduced image is also smaller than the
observation.  For the image in Fig.~2, we use parameters given by KC:
$L_{w} = 5 \times 10^{38}$ erg s$^{-1}$, $R_s = 3 \times 10^{17}$ cm,
$\gamma_w = 3 \times 10^6$, and $p = 3$. In this case, we have $\nu
L_\nu \sim 10^{36}$ erg~sec$^{-1}$ at 1 keV.

\section{Discussion}

Applying the KC model, we reconstruct an X-ray image which is found to
be inconsistent with the Chandra image.  Owing to a pure toroidal field
and uniform pitch angle distribution, the reproduced image is not
ring-like but 'lip-shaped'. Furthermore, the surface brightness contrast
between the front and back sides of the ring is much less than the
observed. The weak contrast is simply due to the smallness of $\sigma$,
by which the postshock flow slows down quickly after the shock.  The
assumptions of the toroidal field and the smallness of $\sigma$ are thus
found to be incompatible with the observation.

If we assume isotropic emission in the proper frame such as is expected
in a turbulent field, then the ring-like structure is reproduced as
shown in Fig.~4.  As would be expected, we find that such a turbulent
component must be at least comparable to the mean toroidal field in
order to reproduce the ring image. Although an another solution can be
to adopt a contrived pitch angle distribution, we think this is
unlikely.  The image in Fig.~4 is produced in the following way: (1)
assume the magnetic field is random so that the emissivity is isotropic
in the proper frame; (2) set the flow velocity to be 0.2$c$ by hand,
ignoring the flow dynamics; (3) let the distribution function and the
field strength follow the KC model.  Thus, the random field and the fast
flow are essential to reproduce the image.
 
\begin{figure}
  \begin{center}
    \includegraphics[width=15pc]{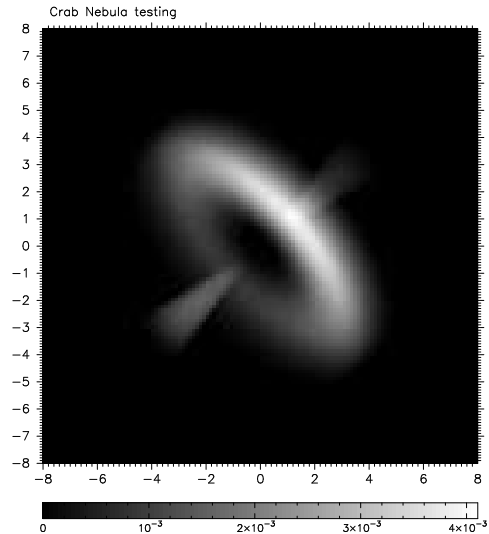} 
  \end{center}
  \caption{An image reproduced with assumptions of
a turbulent field and a high speed flow.
See text in detail.}
\end{figure}

With this practice, we suggest that the nebula field is far from pure
toroidal, but rather is disordered.  Such a disordered field can be
produced by magnetic reconnection or some instability of the toroidal
flux tubes. If there is dissipation of the magnetic field, the flow
dynamics is importantly changed, as is the flow speed.
Even if the value of $\sigma$, which is defined in the wind, is not small,
dissipation in the nebula flow causes deceleration and
brightening. In this sence, $\sigma$ is effectively small so that
the luminosity of the nebula will be explained as the KC model.
But, such a simple heating may not always be good for explaining 
the surface brightness contrast because of the deceleration.
Recently, Komissarov \& Lyubarsky (2003) provide an MHD simulation for
the Crab Nebula, suggesting a complicated flow pattern and
a high speed flow such that the brightness contrast can 
be reproduced. Three dimensional motions associated with magnetic 
energy conversion in the nebula will
considerably change the picture of the nebula.
Numerical simulations for the nebular flow
is of particluar importance in the future study.
As noted, the present image-production scheme will be easily extended 
to combine with such mumerical results.

A model explaining the Chandra observation may be constructed if we
assume a larger $\sigma$ and a subsequent magnetic energy conversion
into heat and plasma kinetic energy, such as magnetic reconnection,
in the nebula flow. Suppose $\sigma$ is rather
large, then the postshock flow must be faster. The inner ring is formed
at the shock. The brightness distribution will be similar to that of
$\sigma = 0.1$ in Fig.~3.  As the flow proceeds outward, the magnetic
energy conversion takes place (accelerating and heating the flow). 
This causes the second brightening. 
Subsequently, the synchrotron burn-off provides the outer
boundary of the torus. The Doppler effect will cause a higher brightness
contrast.  We note that the smallness of $\sigma$ is not obvious if
non-ideal-MHD is introduced in the nebula flow.

The above hypothesis explains the disk formation.  Obliqueness of the
pulsar causes a series of current sheets with an interval of the light
cylinder radius ($\sim 10^8$cm) in the equatorial region.  If the
current sheets dissipate in the nebula, the synchrotron emission
brightening is restricted in the equatorial region, where reconnection
takes place.  The appearance of pulsar nebulae should depend on
obliqueness of individual pulsars. High obliqueness results in a thick
disk and a high efficiency of synchrotron luminosity, while
near-alignment causes a faint nebulae.

The possibility of a dissipative process in the nebula may be examined
more rigorously with spatially-resolved spectra, for which we will
compare the model and the observation in a subsequent paper.

\section*{Acknowledgments}
The authors thank F. E. Bauer for careful review of the manuscript. This
work was supported by JSPS KAKENHI (C), 15540227.  K. M. is supported by
the JSPS postdoctoral fellowship for research abroad.

\label{lastpage}
\end{document}